\documentclass[Entropy,article,accept,moreauthors,dvipdfm,12pt,a4paper]{mdpi} 
\lastpage{1854}
\doinum{10.3390/e12071833}
\pubvolume{12}
\pubyear{2010}
\history{Received: 24 June 2010; in revised form: 20 July 2010 / Accepted: 21 July 2010 / \linebreak Published: 23 July 2010}

\usepackage{caption}
\usepackage{hyphenat}
\usepackage[sort&compress]{natbib}
\usepackage{hypernat}

\usepackage{amsmath,amsthm,amsfonts,amscd,eucal,latexsym,amssymb}
\Title{Black Hole Horizons and Thermodynamics: A Quantum Approach}

\Author{Valter Moretti $^{1,\star}$ and Nicola Pinamonti $^{2}$}

\address{%
$^{1}$ Dipartimento di Matematica, Universit\`a di Trento
 and  Istituto Nazionale di Fisica Nucleare--Gruppo Collegato di Trento, via Sommarive 14
I-38050 Povo (TN), Italy\\
$^{2}$ Dipartimento di Matematica, Universit\`a di Roma ``Tor Vergata", Via della Ricerca Scientifica 1, I-00133 Roma, Italy;
E-Mail: pinamont@mat.uniroma2.it}

\corres{E-Mail: moretti@science.unitn.it.}

\abstract{We focus on quantization of the metric of a black hole
restricted to the Killing horizon with universal radius $r_0$. After
imposing spherical symmetry and after restriction to the Killing
horizon, the metric is quantized employing the chiral currents
formalism. Two ``components of the metric'' are indeed quantized:
The former behaves as an affine scalar field under changes of
coordinates,   the latter is instead a proper scalar field. The
action of the symplectic group on both fields is realized in terms
of certain horizon diffeomorphisms. Depending on the choice of the
vacuum state, such a representation is  unitary. If the reference
state of the scalar field is a coherent state rather than a vacuum,
spontaneous breaking of conformal symmetry arises and the state
contains a Bose-Einstein condensate. In this case the order
parameter fixes the actual size of the black hole with respect to
$r_0$. Both the constructed state together with the one associated
with the affine scalar are thermal states (KMS) with respect to
Schwarzschild Killing time when restricted to half horizon. The
value of the order parameter fixes the temperature at the Hawking
value as well. As a result, it is found that the  quantum  energy
and entropy densities coincide with the black hole mass and entropy,
provided the universal parameter $r_0$ is suitably chosen, not
depending on the size of the actual black hole in particular.}

\keyword{black hole; conformal field theory; quantum gravity}

\def\spa{\hskip -3pt}

\def\cL{{\ca L}}

\def\cH{{\ca H}}
\def\cD{{\ca D}}

\def\cT{{\bf T}}

\def\cC{{\ca C}}
\def\cX{{\ca X}}

\def\cV{{\ca V}}
\def\cK{{\ca K}}

\def\bC{{\mathbb C}}           

\def\bN{{\mathbb N}}
\def\bM{{\mathbb M}}

\def\bR{{\mathbb R}}
\def\bP{{\mathbb P}}
 \def\bF{{\mathbb F}}
\def\bS{{\mathbb S}}

\def\bZ{{\mathbb Z}}

\def\gF{{\mathfrak F}}

\def\beq{\begin{eqnarray}}
\def\eeq{\end{eqnarray}}
\def\pa{\partial}
\def\at{\left(}               
\def\ag{\left\{}              

\def\ct{\right)}              
\def\cg{\right\}}             
\newcommand{\ca}[1]{{\cal #1}}         

\def\be{\beta}
\def\ga{\gamma}
\def\de{\delta}

\def\Ga{\Gamma}

\def\La{\Lambda}

\def\trho{\widetilde{\rho}}
\def\tX{\widetilde{X}}

\newcommand{\media}[1]{\langle{#1}\rangle}

\begin{document}

\section{Introduction and Summary}
After the work of Bekenstein and Hawking \cite{beke72, beke73}
defining the thermodynamical properties of black holes
\cite{hawk75,P05},  microscopic explanations of thermodynamical
features of black holes became a standard issue of modern
theoretical physics. A microscopic explanation should throw some
light on a possible quantum description of gravity. The holographic
principle, proposed by 't Hooft and Susskind
\cite{thoo93,thoo95,suss95,P04}, suggests to search exactly on event
horizons the quantum nature of the gravity. That is a reason why
many scientists in the last decades have tackled the problem under
different points of view obtaining some relevant but partial results
\cite{Wald99,Furs04,P05}.

Restricting to the case of spherical symmetry, we propose here an
alternate approach, regarding  canonical quantization (by means of a
straightforward extension of the notion of chiral current) of some
``components'' of the metric when  restricted to the horizon of a
black hole. This approach is based on  results previously obtained
by the authors \cite{MP6,MP66}, using some of the ideas of \cite{GLRV},
where the language and the mathematical tools of conformal nets of
local observables were exploited profitably. It is worth remarking
that the approach employed in this paper has recently been developed
toward other very promising directions in quantum field theories in
curved spacetime \cite{DMP,Moretti06,Moretti08,DMP2,DPP0}. Herein,
instead, we focus to quantum gravity. Our main idea is that the
metric of a black hole, restricted to the Killing horizon with
radius $r_0$ and represented in a suitable manner after imposing
spherical symmetry, can be quantized with the procedures of chiral
currents. We quantize two ``components of the metric'': one behaves
as an affine scalar fields under changes of coordinates and the
other is a proper scalar field. However, the reference state of the
latter, which, in fact fixes the black hole, is not a vacuum state
but a coherent state arising from spontaneous breaking of conformal
symmetry (which is a natural symmetry on null surfaces). That state
(and the state associated with the affine field), when examined on a
half horizon, turns out to be thermal (KMS) with respect to
Schwarzschild Killing time (restricted to the horizon) and it
contains a Bose-Einstein condensate. The value of order parameter
individuates the actual size  of the black hole with respect to
$r_0$, as well as the Hawking temperature. In spite of initial
limitation to only treat the spherical case, we finally find that
the densities, energy and entropy of this state coincide to the mass
and the entropy of the black hole provided the  universal parameter
$r_0$ is fixed suitably (not depending on the size of the black
hole).

Let us review some basic results of the theory of $n$-dimensional
($n\geq 3$) vacuum Einstein solution (with cosmological constant
$\Lambda$) enjoying spherical symmetry. Such metrics have the
following general form in a coordinate patch where $\eta>0$
\beq\label{metric} ds^2=\frac{g_{ab}(x)}{\eta(x)^{\frac{n-3}{n-2}}}
\: dx^adx^b+\eta(x)^{\frac{2}{n-2}} \:d\Sigma^2(y)\:. \eeq Above
$a,b=0,1$, moreover the ``non-angular part of the metric'' is given
by the Lorentzian metric $g$ (with signature $-,+$) which, together
with the so-called {\em dilaton field} $\eta$, depends only on
coordinates $x^a$. The ``angular part of the metric'' $d\Sigma^2 =
h_{AB} dy^A dy^B$ with $A,B=1,2,\cdots, n-2$, depends on coordinates
$y^A$ only.  $h$ is the metric on a Riemannian  $(n-2)-$dimensional
space $\Sigma$ and it is supposed to satisfy, if $R_{AB}[h]$ is
Ricci tensor associated with $h$, \beq R_{AB}[h]= \Gamma \:
h_{AB}\:,\:\:\:\: \mbox{with $\Gamma$ constant.} \label{added}\eeq
Under these assumptions  vacuum Einstein equations give rise to
equations for the metric $g$ and the dilaton $\eta$ which can also
be obtained by means of a variational principle. Starting from
Hilbert-Einstein action for the complete metric (\ref{metric}) and
integrating out the angular part discarding it, one obtains:
\beq\label{2daction} I[g,\eta]=\frac{2}{G}\int dx^0dx^1\:\sqrt{|det
g|}\: \left\{ \eta\frac{R[g]}{2}+\cV(\eta)\right\}\:. \eeq That is
the action of a $2$-D dilatonic theory with dilatonic potential \beq
\cV(\eta)=\frac{\Ga}{2\eta^{\frac{1}{n-2}}}-\La\,{\eta^{\frac{1}{n-2}}}\:.\label{V}\eeq
$\cV$ encodes all information about the $n$-dimensional original
spacetime with cosmological constant $\La$. The approach to Einstein
equation based on the action (\ref{2daction}) is called {\em
dimensional reduction} \cite{NO,CU,GKV,Gia,Schmidt}. We stress that,
in spite of the reduction, the finally obtained  $2$-dimensional
models share many properties with $n-$dimensional ones.

We can reduce the number of used fields by the following remark. It
is well known that every Lorentzian $2$-dimensional metric $g$ is,
at least locally, conformally equivalent to the flat metric
\linebreak $\ga = diag(-1,1)$ referred to Minkowski coordinates
$x^0,x^1$: \beq g_{ab}(x)=e^{\rho(x)}\ga_{ab}(x)\:. \label{gamma}
\eeq (In several papers the exponent is defined as $-2\rho$ instead
of $\rho$). Working with null coordinates \linebreak $x^\pm=x^0\pm
x^1$, so that $\gamma_{\pm}= \gamma_{\mp} = -1/2$ and $\gamma_{++}=
\gamma_{--}=0$, one has \beq g_{++}= g_{--} =0\:, \:\:\:\: g_{+-} =
g_{-+} = -e^{\rho(x)}/2\:, \label{gamma2} \eeq and vacuum Einstein
equations become very simple:
\beq\label{eqmoto} \pa_+\pa_-\eta+e^{\rho}\,
\frac{\cV(\eta)}{2}=0,\qquad \pa_\pm^{2}\eta
-\pa_\pm\rho\,\pa_\pm\eta=0,\qquad \pa_+\pa_-\rho+\frac{e^{\rho}}{2}
\frac{d \cV(\eta)}{d\eta}=0\:. \eeq
Starting from the action above
as a functional of the fields $\eta, \rho$ the variational procedure
produces only two equations, the first and the last in
(\ref{eqmoto}), of the original four Einstein equations, the
remaining ones can be imposed as constraints on the solutions.
Equations (\ref{eqmoto}) are completely
integrable and the general solution depends on an arbitrary real
function $\phi(x) =\phi^+(x^+)+\phi^-(x^-)$ and an arbitrary real
\linebreak constant $C$: \beq e^{\rho}=-\frac{F_C(\eta)}{2}\pa_+\phi\pa_-\phi,
\qquad 2G_C(\eta)=\phi\:, \label{one} \eeq where $\eta$ is assumed
to satisfy $\eta >0$ and \beq F_C(\eta) =\int^\eta_0 \cV(\alpha)\;
d\alpha - C,\qquad G_C(\eta) =\int \frac{1}{F_C(\eta)}\; d\eta\:.
\label{two} \eeq The integration constant of the latter integral is
included in the field $\phi$ in (\ref{one}). The explicit expression
for $F_C$ reads \beq F_C(\eta) =\at\frac{\Ga}{2}\frac{n-2}{n-3}-\La
\frac{n-2}{n-1}\eta^{\frac{2}{n-2}}\ct\eta^{\frac{n-3}{n-2}}-C
\:\:\:\:\mbox{or $\:\:F_C(\eta) = -\frac{\Lambda \eta^2}{2} - C\:\:$
if $n=3$.} \eeq With these definitions the metric (\ref{metric})
takes the following explicit form \beq\label{2dschmet0}
ds^2=\frac{F_C(\eta(\phi))}{2\;\eta(\phi)^{\frac{n-3}{n-2}}}\partial_+\phi
\partial_-\phi \:\: dx^+dx^-+\eta(\phi)^{\frac{2}{n-2}} d\Sigma^2\:.
\eeq In $I_+\times I_- \times \Sigma$ ($I_\pm$ being any pair of
(open) segments where $\partial_\pm \phi \neq 0$,
$F_C(\eta(\phi))\neq 0$) the fields $\phi^\pm$ together with
coordinates on $\Sigma$ define a coordinate patch of the spacetime
where \beq\label{2dschmet}
ds^2=\frac{F_C(\eta(\phi))}{2\;\eta(\phi)^{\frac{n-3}{n-2}}}\:d\phi^+
d\phi^- +\eta(\phi)^{\frac{2}{n-2}}  d\Sigma^2 \:. \eeq As  the
metric depends only on  $\phi^++\phi^-$, $\pa_{\phi^+}-\pa_{\phi^-}$
is a {\em Killing  field}. This is a straightforward generalization
of Birkhoff theorem \cite{Schmidt}. The arbitrariness, due to an
additive constant, in defining $\phi^+$ and $\phi^-$ from $\phi$
does not affect the Killing field and it reduces to the usual
arbitrariness of the origin of its integral curves.

As a comment, notice that in the three dimensional case ($n=3$), the
last equation in (\ref{eqmoto}) becomes the field equation of a $2D$
{\em Liouville theory} for $\varrho = \rho/2$ with $k = -\Lambda/4
>0$ \beq -\partial_+\partial_-\varrho + k e^{2\varrho} = 0\:.
\label{liouville} \eeq Notice however that the action in
(\ref{2daction}) does not reduce to the usual Liouville action  in
this case. Restricting to the case  $n=4$ with $\La=0$ and $\Gamma =
2$, two relevant cases arise. If  $C>0$,  the metric
(\ref{2dschmet})  is {\em Schwarzschild's} one  with black-hole mass
$M = C/4$, $r=\sqrt{\eta}$, $\phi/2$ is the  ``Regge-Wheeler
tortoise coordinate'' $r_*$ and $\phi^\pm$  are the usual null
coordinates. As $F_C$ has a unique non-integrable zero,  there are
two inequivalent functions $G_C$ corresponding to the internal
singular metric and the external static metric respectively.
$\pa_{\phi^+}-\pa_{\phi^-}$ defines Schwarzschild time in the
external region. The case $C=0$ is nothing but  Minkowski spacetime.
$\pa_{\phi^+}-\pa_{\phi^-}$ is the Killing field associated with
Minkowski time, there is a unique function $G_C$, and coordinates
$\phi^\pm$ are the usual global (radial) null coordinates with range $\phi^++\phi^->0$.\\
Global structures are constructed gluing together solutions of
Einstein equations.  In particular,  manifolds with {\em bifurcate
Killing horizon} arise. Consider again  the case $n=4$, $C= 4M>0$,
$\La=0$, $\Ga=2$ (Schwarzschild black hole). In this cases one
fixes global coordinates $X^+ \in \bR$, $X^-\in \bR$ such that the
metric reduces to (\ref{2dschmet}) in each of the four sectors
$X^+\lessgtr 0, X^-\lessgtr 0$. $\phi$, $\rho$ and $\eta$ are
functions of $X^\pm$ defined as follows: \beq \phi(X^+,X^-) = 4M
\left(1+ \ln \left|\frac{X^+X^-}{32 M^3}\right|\right),\:\:\:
\rho(X^+,X^-) = 1-\frac{\sqrt{\eta(X^+,X^-)}}{2M}\:,\label{phi} \eeq
 $\eta$ is obtained
by solving, for $0< \eta< (2M)^2$ and $\eta> (2M)^2$ respectively, the equation
\beq
 \sqrt{\eta(X^+,X^-)} + 2M \ln\left| \frac{\sqrt{\eta(X^+,X^-)}}{2M}-1\right| = \frac{\phi(X^+,X^-)}{2} \label{phi0}.
\eeq The global metric (\ref{metric}) obtained in this way (with
$x=X$) is smooth for $\eta(X^+,X^-)>0$, $\eta=0$ being the
black/white-hole singularity. The spacetime obtained  is maximally
extended and $K = \pa_{\phi^+}-\pa_{\phi^-}$ turns out to be a
Killing field smoothly defined globally which is light-like on a
pair of $3$-dimensional null hypersurfaces $\bF$ and  $\bP$. These
hypersurfaces intersect at the compact $2$-dimensional spacelike
submanifolds $\Sigma$, localized at $X^\pm_\Sigma=0$,  and are
normal to it. $\Sigma$ is called {\em bifurcation surface}. $K$
vanishes exactly at the bifurcation surface. It turns out that
$\eta|_{\bP} = \eta|_{\bF} = \eta_C>0$ is  {\em constant} and it is
the {\em unique positive solution} of $F_C(\eta)=0$.
$\sqrt{\eta_C}=2M$ is the  Schwarzschild radius. Either $\bF$ and
$\bP$  are  diffeomorphic to $\bR\times \Sigma$ with $\bR$ covered
by the coordinate $X^+,X^-\in \bR$ respectively. In coordinates
$X^\pm$ it holds: \beq \rho|_\bF = \rho|_\bP=  0\:. \label{1/2}\eeq
Notice that $\phi,\eta,\rho$ depend on the product $X^+X^-$ only.
Therefore, passing to new  global coordinates  $X'^\pm = C_\pm
X^\pm$ with constants $C_\pm$ satisfying $C_+C_- =1$, Equations
(\ref{phi})--(\ref{1/2}) and $X^\pm_\Sigma=0$ still
hold  for the considered metric replacing $X^\pm$ with $X'^\pm$.

The point of view we wish to put forward is to promote (some of) the
objects $\rho$, $\eta$ and $\phi$ to  quantum objects, {\em i.e.},
averaged values of associated  quantum fields $\hat\rho$, $\hat\eta$
and $\hat\phi$ with respect to  reference quantum states. Those
quantum states do individuate the actual metric  and in particular
the mass of the black hole on a hand. On the other hand they should
account for thermodynamical properties of black holes.

In the rest of the paper we adopt of Planck units so that
$\hbar=c=G=k_B=1$, in this way every physical quantity is a pure
number.

\section{Quantum Structures on the Horizon of a Black Holes}
\subsection{Geometrical Background and its Quantum Interpretation}
To go on with our proposal we have to consider as separated objects
part of the background manifold (not quantized) and part of the
metric structure (at least partially quantized). More precisely we
consider a $4$-dimensional differentiable manifold $\bM$
diffeomorphic to $(\bR\times \Sigma)\times (\bR\times \Sigma) =
\bR^2\times \Sigma$ such that a reference flat Lorentzian $2D$
metric $\gamma$ is assigned in a global coordinate frame $(x^+, x^-)
\in \bR\times \bR$ where the two factors $\bR$ are those in the
decomposition of $\bM=(\bR\times \Sigma)\times (\bR\times \Sigma)$.
These coordinates, together with coordinates on $\Sigma$, describe
respectively manifolds $\bF$ and $\bP$. Every admissible metric on
$\bM$ must be such that, {\bf (C1)} it has the general structure
(\ref{metric}) and in particular it enjoys $\bS^2$-spherical
symmetry, $\Sigma$ being tangent to the associated Killing fields,
{\bf (C2)} it solves  equations (\ref{eqmoto}) with $C>0$ (that is
the mass $M$) fixed {\em a priori} in some way depending, at
quantum level, on a quantum reference state as we shall discuss
shortly, and {\bf (C3)}  $\bF \cup \bP$ is a bifurcate Killing horizon with bifurcation surface $\Sigma$.\\
A time orientation is also assumed for convenience by selecting one
of the two disjoint parts of $\bF\setminus \Sigma$ and calling it
$\bF_>$. The other will be denoted by $\bF_<$. (In \cite{MP6,MP66}
we used notations $\bF_\pm$ instead of $\bF_\gtrless$.)

There is quite a large freedom in choosing global coordinates
$x^\pm\in \bR$  on $\bF,\bP$ respectively, such that the form of the
metric (\ref{gamma2}) hold. In the following we call {\bf admissible
null global frames} those coordinate frames. It is simply proved
that the following is the most general transformation between pairs
of admissible null global frames provided  $\eta$ transforms as a
scalar field (as we assume henceforth): \beq \label{transformations}
{x'}^+ = f_+(x^+)\:, \:\:\:\: {x'}^-= f_-(x^-) \:\:\:   \:\mbox{and
$\:\:\:\frac{df_+}{dx^+}\frac{df_-}{dx^-}>0$} \:, \eeq where the
ranges of the functions $f_\pm$ cover the whole real axis.
 We remark that preservation of the form of the metric (\ref{gamma2})
entails preservation of the form of  equations (\ref{eqmoto}).
There are infinitely many possibilities to assign the metric fulfilling the constraints (C1), (C2), (C3).
 Considering Kruskal spacetime, if  $X^\pm_\Sigma=0$
and if $f_\pm: \bR \to \bR$ are functions as in
(\ref{transformations}) with $f_\pm(0)=0$, the fields  $\phi'(X) =4M
\left(1+ \ln \left|\frac{f_+(X^+)f_-(X^-)}{16 M^2}\right|\right)$
give rise to everywhere well-defined fields $\eta$ and $\rho$ using
(\ref{one}) and (\ref{two}). The  produced  spacetime has a
bifurcate Killing horizon with respect to the  Killing vector
$\pa_{\phi'^+}-\pa_{\phi'^-}$ (which has the  temporal orientation
of $\pa_{\phi^+}-\pa_{\phi^-}$) just determined by the initially
assigned manifolds $\bF,\bP,\Sigma$. Different global metrics
obtained from different choices of the functions $f_\pm$ are however
diffeomorphic, since they have the form of Kruskal-like metric (with
the same mass) in admissible null global frame, $x^\pm=
f_\pm(X^\pm)$ in the considered case.

\subsection{The Field $\rho$ and the Interplay with $\phi$ on Killing Horizons}\label{sec1}
As a consequence of the decomposition $\phi(x) =
\phi^+(x^+)+\phi^-(x^-)$,  $\phi$ is a solution of d'Alembert
equation $\Box \phi=0$, where $\Box$ is referred to the reference
flat metric $\gamma$ in any admissible null global frame. This field
is a good candidate to start with a quantization procedure. In
particular each component $\phi^\pm(x^\pm)$ of $\phi$ could be
viewed as a scalar quantum field on $\bF$ and $\bP$ respectively.
Here we focus also attention of the field $\rho$ and on the
interplay between $\rho$ and $\phi$ when restricted to
 the horizon.

Let us start by showing {\em classical} nontrivial properties of the
field $\rho$ and its restrictions $\rho|_\bF$ and $\rho|_\bP$. First
of all consider transformations of coordinates
(\ref{transformations}) where, in general,  we relax the requirement
that coordinates $x'^+,x'^-$ are global and we admit that the ranges
of functions $f_\pm$ may be finite intervals in $\bR$. The field
$\rho$ transforms as \beq \rho(x'^+,x'^-) = \rho(x^+(x'^+),
x^-(x'^-)) + \ln \frac{\partial (x^+,x^-)}{\partial(x'^+,x'^-)}\:,
\label{affinerho} \eeq where the argument of $\ln$ is  the Jacobian
determinant of a transformation $x=x(x')$. (\ref{affinerho}) says
that the field $\rho$ transform as an {\em affine scalar} under
changes of coordinates. (We notice {\em en passant} that, from
(\ref{one}) and (\ref{two}) and the fact that  $\eta$ is a scalar
field, (\ref{affinerho}) entails that $\phi$ is  a scalar field as
assumed previously.) A reason for the affine transformation rule
(\ref{affinerho}) is that, for the metric $g$, only Christoffel
symbols $\Ga_{++}^+$ and $\Ga_{--}^-$ are non vanishing and \beq
\pa_+ \rho = \:\Ga_{++}^+\:,\:\:\:\:\pa_- \rho =\Ga_{--}^-\:. \eeq
{\bf Remark}. The reader should pay attention to the used notation.
$\rho$ should be viewed as a function of both the points of
$\bF\times \bP$ and the used chart. If the chart $\cC$ is associated
with the coordinate frame $x^+,x^-$, an appropriate notation to
indicate the function representing $\rho$ in $\cC$ could  be
$\rho(\cC|x^+,x^-)$ or $\rho_\cC(x^+,x^-)$. However we shall use the
simpler, but a bit miss-understandable, notation $\rho(x^+,x^-)$. As
a consequence, the reader should bear in his/her mind that, in
general
$$\rho(x'^+,x'^-) \neq \rho(x^+(x'^+), x^-(x'^-))\:.$$

Form a classical point of view, on a hand $\rho|_\bF$ and
$\rho|_\bP$ embody all information about the metric, since they
determine completely it via Einstein equations, on the other hand
these restrictions can be  assigned freely. More precisely the following  theorem  holds.\pagebreak

\noindent {\bf Theorem 1}. {\em Working in a fixed admissible null global frame $x^\pm$ on $\bM$,
if $\rho_+=\rho_+(x^+)$, $\rho_-=\rho_-(x^-)$ are smooth bounded-below functions,
there is a unique metric which satisfies (C1), (C2), (C3) (with assigned mass $M>0$) and such that  $\rho|_\bF= \rho_+$
and $\rho|_\bP= \rho_-$}.\\

\noindent {\bf Proof}. First of all we prove that if  $ds^2$,
$\widetilde{ds^2}$ are solutions of Einstein equation on $\bM$
satisfying (C1), (C2), (C3) (with a fixed value of the mass), they
coincide if the restrictions of the respectively associated
functions  $\rho$, $\trho$ to $\bF\cup \bP$ coincide working in some
admissible null global frame $x^\pm$.

Indeed, using (\ref{affinerho}) one sees that if $\rho|_\bF=
\trho|_\bF$ and $\rho|_\bP= \trho|_\bP$ in an admissible null global
frame, these relations must hold in any other admissible null global
frame. Hence we consider, for the metric $ds^2$, the special
admissible null global frame  $X^\pm$ introduced in the end of the
introduction. In these coordinates it must hold $\rho|_\bF =
\trho|_\bF = \rho|_\bP = \trho|_\bP = 0$. On the other hand,
$\trho|_\bF = \trho|_\bF =0$ is valid also in coordinates $\tX^\pm$
analog of $X^\pm$ for the metric $\widetilde{ds^2}$. Applying
(\ref{affinerho}) for $\trho$ with respect to the coordinate systems
$X^\pm$ and $\tX^\pm$ one easily finds (assuming that
$\tX^\pm_\Sigma= X^\pm_\Sigma=0$ for convenience) $X^\pm= C_\pm
\tX^\pm$ such that the constants $C_\pm$ satisfy $C_+C_-=1$.
Therefore, in coordinates $X^\pm$, $\widetilde{ds^2}$ has the form
individuated by (\ref{phi}) and (\ref{phi0})  and thus it coincides
with $ds^2$. This facts is invariant under transformations
(\ref{affinerho}) and so, in particular, the affine scalars $\rho$
and $\rho'$ coincide also in the initial reference frame. The $n=3$
case is analog.

To conclude, let us prove the existence of a metric satisfying (C1),
(C2), (C3) when restrictions of $\rho$ to $\bF\cup \bP$ are assigned
in a global null admissible coordinate frame. In the given
hypotheses, by direct inspection one may build up a global
transformation of coordinates $x^\pm \to X^\pm$ as in
(\ref{transformations}), such that $\rho_+(X^+)=\rho_-(X^-)=0$
constantly. Now a well-defined metric compatible with the bifurcate
Killing horizon structure can be defined  as in (\ref{phi}). This
metric is such that $\rho$ reduces to $\rho_+(X^+)$ on $\bF$ and
$\rho_-(X^-)$ on $\bP$. Transforming back everything in the initial
reference frame $x^\pm$, the condition $\rho|_\bF = \rho_+$,
$\rho|_\bP = \rho_-$ turns out to be preserved  trivially
by transformations (\ref{transformations}). The $n=3$ case is analog. ~~~~~~~~~$\Box$ \\

Working in a fixed admissible global null coordinate frame, consider
the restriction of $\rho$ to $\bF$: $\rho|_\bF(x^+) =
\rho(x^+,x^-_\Sigma)$. The transformation rule of $\rho|_\bF(x^+)$
under changes of coordinates makes sense only if we consider a
change of coordinates involving both $x^+$ and $x^-$. It is not
possible to say how $\rho(x^+)$ transforms  if only the
transformation rule $x'^+ = f_+(x^+)$  is known whereas $x'^- =
f_-(x^-)$ is not. This is because (\ref{affinerho}) entails \beq
\label{relevant}\rho(x'^+, x'^-_\Sigma) = \rho(x^+, x^-_\Sigma) +\ln
\frac{dx^+}{dx'^+} + \left.\ln
\frac{dx^-}{dx'^-}\right|_{x^-_\Sigma}\:.\eeq However, since the
last term in the right-hand side is constant on $\bF$, the
transformation rule for field $\partial_{x^+}\rho(x_+, x^-_\Sigma)$
is well-defined for changes of coordinates in $\bF$ only $x'^+ =
f_+(x^+)$. This situation resembles that of $\phi$. The restriction
of $\phi$ to $\bF$ is ill defined due to the divergence of
$\phi^-(x^-)$ at $x^-_\Sigma$ (see (\ref{phi})), whereas the
restriction of $\partial_+\phi$ is well-defined  and it coincides
$\partial_+\phi^+$, the arbitrary additive constant in the
definition of $\phi^+$ being not relevant due to the presence of the
derivative. The analogs  hold  replacing $\bF$ with $\bP$.

\subsection{Quantization}
Quantization of $\rho$ and $\phi$ in the whole spacetime would
require a full quantum interpretation of Einstein equations, we
shall not try to study that very difficult issue. Instead, we
quantize $\rho|_\bF$  (actually the derivatives of that field)
which, classically, contains the full information of the metric but
they  are not constrained by any field equations. Similarly we
quantize $\phi|_\bF = \phi^+$ (actually its derivative) and we
require that classical constraints  hold for its mean value (with
respect to that of $\hat\rho$), which is required  to coincide with
the classical field $\phi^+$ (actually its derivative). We show
that, in fact there are quantum states which fulfill this constraint
and enjoy  very interesting physical properties.

 From now on we consider the quantization procedure for  fields
$\hat{\rho}_\bF$ and $\hat{\phi}^+$ defined on the metrically
degenerate hypersurface $\bF$. Since we consider only quantization
on $\bF$ and not on $\bP$, for notational simplicity we omit the the
indices  $_\bF$ and $^+$ of $\hat{\rho}_\bF$ and $\hat{\phi}^+$
respectively and we write $\hat{\rho}$ and $\hat{\phi}$ simply.
Omitting complicated mathematical details, we adopt  canonical
quantization procedure on null manifolds introduced in
\cite{MP2,MP4} and developed in \cite{MP6,MP66} for a real scalar field
as $\phi$ based on Weyl algebra. This procedure gives rise to a nice
interplay with conformal invariance studied in various contexts
\cite{MP1,MP2,MP4,MP6,MP66}.

It is convenient to assume that $\hat\rho$ and $\hat\phi$ are
function of $x^+$ but also of angular coordinates $s$ on $\Sigma$:
{\em The Independence from angular coordinates will be imposed at
quantum level picking out a $2$-dimensional spherically symmetric
reference state}. $\Sigma$ is supposed to be equipped with the
metric of the $2$-sphere with radius $r_0$, it being  a universal
number to be fixed later. Notice that, as a consequence $r_0$ does
not depend on the mass of any possible black hole. A black hole is
selected by fixing a quantum state.

We assume that only transformations of coordinates which do not mix
angular coordinates $s$ and  coordinate $x^\pm$ are admissible.
Under transformations of angular coordinates, $s'=s'(s)$,
$\partial_{x^+}\hat\rho$ transform as a usual scalar field, whereas
it transforms as a connection symbol under transformations of
coordinates $x'^+= x'^+(x^+)$ with positive derivative: \beq
\partial_{x'^+}\hat{\rho}(x'^+,s') =
\frac{dx^+}{dx'^+}\: \partial_{x^+}\hat{\rho}(x^+,s) +
\frac{dx'^+}{dx^+}\:\frac{d^2x^+}{{dx'^+}^2}I\:. \label{affinerhoF}
\eeq Conversely $\hat\phi$ transforms as a proper scalar field  in
both cases with the consequent transformation rule for
$\partial_+\hat\phi$: \beq
\partial_{x'^+} \hat{\phi}(x'^+,s') =
\frac{dx^+}{dx'^+}\: \partial_{x^+}\hat{\phi}(x^+,s)
\label{affinerhoFbis}\:. \eeq Transformation rules for the field
$\hat \rho$ are not completely determined from (\ref{affinerhoF}).
However, as explained below it does not matter since the relevant
object is $\partial_+\hat\rho$ either from a physical and
mathematical point of view.

Fix an admissible (future oriented for convenience) null global
frame $(V, s)$ on  $\bF= \bR\times \Sigma$. For sake of simplicity
we assume that $V_\Sigma=0$ so that the bifurcation surface is
localized at the origin of the coordinate $V$ on $\bR$. In
coordinate $(V,s)$  Fock representations  of $\hat \phi$ and
$\hat\rho$ are obtained as follows  \cite{MP6,MP66} in terms of a
straightforward generalization of chiral currents
 (from now on, $n=0, \pm 1,\pm 2,\cdots$ and $j= 1,2,\cdots$ and where $\theta(V)= 2\tan^{-1}V$):
\begin{eqnarray}
\hat{\phi}(V,s) &=&\frac{1}{i\sqrt{4\pi}}\sum_{n,j} \frac{u_j(s)e^{-in\theta(V)}}{n}J^{(j)}_{n}\:, \label{fieldphi}\\
\hat{\rho}(V,s) &=&\frac{1}{i\sqrt{4\pi}}\sum_{n,j} \frac{w_j(s)e^{-in\theta(V)}}{n}P^{(j)}_{n}\:. \label{field}
\end{eqnarray}
As $V$ ranges in $\bR$, $\theta(V)$ ranges in $[-\pi,\pi]$. (The
identification $-\pi\equiv \pi$ would make compact the horizon,
which would become $\bS^1 \times \Sigma$,  by adding a point at
infinity to every null geodesic on $\bF$. This possibility will be
exploited shortly in considering the natural action of the conformal
group $PSL(2,\bR)$.) $u_j$ and $w_j$ are {\em real} and, separately,
define  Hilbert bases in $L^2(\Sigma, \omega_\Sigma)$  with measure
$\omega_\Sigma= r_0^2 \sin\vartheta d\vartheta \wedge d\varphi$.
There is no cogent reason to assume $u_j=w_j$ since the results are
largely independent from the choice of that Hilbert basis. Operators
$J^{(j)}_{n}, P^{(j)}_{n}$  are such that $J^{(j)}_{0}=P^{(j)}_{0} =
0$ and
 $J^{(j)\dagger}_{n}= J^{(j)}_{-n}$, $P^{(j)\dagger}_{n}= P^{(j)}_{-n}$
and oscillator commutation relations for two independent systems are valid
\begin{eqnarray}
[J^{(j)}_{n}, J^{(j')}_{n'}] &=& n \delta^{jj'}\delta_{n,-n'}I\:, \label{commutationsphi}\\
{[}P^{(j)}_{n}, P^{(j')}_{n'}{]} &=& n \delta^{jj'}\delta_{n,-n'}I\:, \label{commutations}\\
{[}J^{(j)}_{n}, P^{(j')}_{n'}{]} &=& 0\:.
\end{eqnarray}
The space of the representation is the tensor products of a pair of
bosonic Fock spaces $\gF_\Psi\otimes\gF_\Upsilon$ built upon the
vacuum states $\Psi, \Upsilon$ such that $J^{(j)}_{n} \Psi = 0$,
$P^{(k)}_{m} \Upsilon = 0$ if $n,m\geq 0$, while the states with
finite number of particles are obtained, in the respective Fock
space, by the action of operators $J^{(j)}_{n}$  and  $P^{(k)}_{m}$
on $\Psi$ and $\Upsilon$ respectively for $n,m<0$.

 From a mathematical point of view it is important to say that the
fields $\hat{\phi}(x^+,s)$ and $\hat{\rho}(x^+,s)$ have to be
smeared by integrating the product of $\hat{\phi}(x^+,s)$,
respectively  $\hat{\rho}(x^+,s)$, and a differential form $\omega$
\linebreak of shape
$$\omega = \frac{ \partial f(x^+,s)}{\partial x^+} dx^+\wedge \omega_\Sigma(y)\:,$$ where $f$ is a smooth real scalar
field on $\bF$ compactly supported  and  $\omega_\Sigma$ is the
volume-form on $\Sigma$ defined above. There are several reasons
\cite{MP2,MP4,MP6,MP66} for justify this procedure, in particular the
absence of a measure on  the factor $\bR$ of $\bF= \bR \times
\Sigma$: Notice that forms include a measure to be used to smear
fields, for instance, the smearing procedure for $\hat\phi$ reads
$\int_\bF \hat{\phi}(V,s) \omega(V,s)$ simply. Moreover, this way
gives rise to well-defined quantization procedure based on a
suitable Weyl $C^*$-algebra \cite{MP6,MP66}. Actually, concerning the
field $\hat \rho$ another reason arises from the discussion about
Equation (\ref{relevant}) above. Using $x^+$-derivatives of
compactly supported functions to smear $\hat \rho$ it is practically
equivalent, via integration by parts, to using actually the field
$\partial_+ \hat \rho(x^+,s)$ which is well-defined concerning its
transformation properties under changes of coordinates. Another
consequence of the smearing procedure is the following. Relations
(\ref{commutations}) are equivalent to bosonic commutation relations
for two independent systems
\begin{eqnarray}
[\hat{\phi}(V_1,s_1), \hat{\phi}(V_2,s_2)] &=& \frac{-i}{4}\:\delta(s_1,s_2) \mbox{sign}(V_1-V_2)\: I\:,\\
{[}\hat{\rho}(V_1,s_1), \hat{\rho}(V_2,s_2){]} &=& \frac{-i}{4}\:\delta(s_1,s_2) \mbox{sign}(V_1-V_2)\: I\:,\\
{[}\hat{\phi}(V_1,s_1), \hat{\rho}(V_2,s_2){]} &=& 0\:.
\end{eqnarray}
Indeed, these relations arise
from bosonic quantization procedure based on bosonic Weyl algebra
constructed by a suitable symplectic form, see \cite{MP6,MP66} for full
details.
Actually those relations have to be understood for fields smeared
with forms as said above. Changing coordinates and using
(\ref{affinerhoF}), these relations  are preserved for the field
$\hat{\rho}(x'^+,s')$ smeared with forms since the added term
arising from (\ref{affinerhoF}) is a $\bC$-number and thus it
commutes with operators.

The mean values $\langle \Upsilon | \hat\rho \:\Upsilon \rangle$,
$\langle \Psi | \hat\phi \:\Psi \rangle$ with respect quantum states
$\Upsilon$ and $\Psi$ respectively should correspond (modulo
mathematical technicalities) to the classical function $\rho|_\bF$
and $\phi|_\bF$. Let us examine this case.

By construction $\langle \Upsilon | \hat{\rho}(V,s) \Upsilon
\rangle=0$. This suggest that the interpretation of the coordinate
$V$ must be the global coordinate along the future horizon $X^+$
introduced at the end of the introduction when the mean value of
$\langle \Upsilon| \hat{\rho}(V,s) \Upsilon \rangle$  is interpreted
as the restriction of the classical field $\rho$ to $\bF$. Indeed,
in the coordinate $X^+$, $\rho$ vanishes  on $\bF$. Actually this
interpretation should be weakened because the field must be smeared
with forms to be physically interpreted. In this way $\langle
\Upsilon | \hat{\rho}(V,s) \Upsilon \rangle=0$ has to be interpreted
more properly as $\langle \Upsilon | \partial_V\hat{\rho}(V,s)
\Upsilon \rangle=0$. Thus one cannot say that $V = X^+$ but only
that $V = k X^+$ for some non vanishing constant $k$. Hoverer the
coordinate $X^+$ is defined up to such a transformation provided the
inverse transformation is performed on its companion $X^-$ on $\bP$
(see the end of the introduction). Notice also that by construction
$\rho|_\bF(V) = \langle \Upsilon | \hat{\rho}(V,s) \Upsilon \rangle$
is spherically symmetric since it vanishes. From a semi-classical
point of view at least, one may argue that  the state $\Upsilon$ and
the analog for quantization on $\bP$ referred to a global coordinate
$U$, picks out  a classical metric: It is the metric having the form
determined by Equations (\ref{phi}),(\ref{phi0}) in coordinates $X^+
= V, X^- = U$.

The interpretation of the mean value of $\hat\phi$ is much more
intriguing. Working in coordinates $V$, from the interpretation of
$\langle \Upsilon | \hat{\rho}(V,s) \Upsilon \rangle$ given above
and using  (\ref{phi}) one  expects that the mean value of
$\partial_V\hat \phi(V,s)$ coincides with $\zeta/V$ where $\zeta=
4M$. This is not possible if the reference state is $\Psi$. However,
indicating the field $\hat\phi$ with $\hat{\phi}_\zeta$ for the
reason explained below, as shown in \cite{MP6,MP66} there is a new state
$\Psi_\zeta$ completely defined from the requirement that it is {\em
quasifree} (that is its $n$-point functions are obtained from the
one-point function and  the two-point function via Wick expansion)
and
\begin{eqnarray}
~&~&\langle \Psi_\zeta |\hat \phi_\zeta(V,s) \Psi_\zeta\rangle = \zeta \ln |V|\:,\label{former} \\
\langle \Psi_\zeta |\hat \phi_\zeta(V,s) \hat \phi_\zeta(V',s')\Psi_\zeta\rangle &=& -\frac{\delta(s,s')}{4\pi}\ln|V-V'| +
\zeta^2 \ln |V|\ln|V'
| + R(V) + R(V')\:,\label{latter}
\end{eqnarray}
where the rests $R$ are such that they gives no contribution when
smearing both the fields with forms as said above. In practice,
taking the smearing procedure into account, $\Psi_\zeta$ is the Fock
vacuum state for the new field operator $\hat{\phi}_0$, with \beq
\hat{\phi}_\zeta(V,s) = \hat{\phi}_0(V,s) + \zeta \ln |V|\: I
\label{cl-qu}\:.\eeq Properly speaking the state $\Psi_\zeta$ cannot
belong to $\gF_\Psi$ because, as shown in \cite{MP6,MP66}, $\Psi_\zeta$
gives rise to a non-unitarily equivalent representation of bosonic
commutation relation  with respect to the representation given in
$\gF_\Psi$. For this reason we prefer to use the symbol
$\hat{\phi}_\zeta$ rather than $\hat\phi$ when working with the
representation of CCR based on $\Psi_\zeta$ instead of $\Psi$.
 The picture should be handled in the framework of
{\em algebraic quantum field theory} considering $\Psi_\zeta$ as a {\em coherent state} (see \cite{MP6,MP66} for details).
Notice that (\ref{former}) reproduces the requested, spherically symmetric,  classical value  of $\partial_V\phi|_\bF =
\partial_V\phi^+$.\\

\noindent{\bf Remark}.  Generally speaking, in quantizing gravity
one has to discuss how the covariance under diffeomorphisms is
promoted at the quantum level. In our picture, the relevant class of
diffeomerphisms is restricted to the maps (\ref{transformations}).
Their action at the quantum level is implemented through
(\ref{affinerhoF})  and (\ref{affinerhoFbis}) in terms of a  class
of automorphisms  of the algebra generated by the quantum fields
$\hat{\phi}$ and $\hat{\rho}$. However, introducing quantum states,
the picture  becomes more subtle.  We shall shortly see that  such a
symmetry will be broken, the remaining one being described  by
$PSL(2,\bR)$ or a subgroup.

\subsection{Properties of $\Psi_\zeta$ and $\hat{\phi}_\zeta$: Spontaneous Breaking of
Conformal Symmetry, Hawking Temperature, Bose-Einstein Condensate}
$\Psi_\zeta$ with $\zeta\neq 0$ involves {\em spontaneous breaking
of $PSL(2,\bR)$ symmetry}. This breaking of symmetry enjoys an
interesting physical meaning we go to illustrate. Let us  extend
$\bF$ to the manifold $\bS^1\times \Sigma$ obtained by adding a
point at infinity $\infty$ to every maximally extended light ray
generating the horizon $\bF$. On the circle $\bS^1$ there is a
well-known \cite{MP6,MP66} natural geometric action $PSL(2,\bR) =
SL(2,\bR)/\pm$ (called M\"obius group of the circle) in terms of
diffeomorphisms of the circle. Using global coordinates $V,s$ the
circle $\bS^1$ is parametrized by $\theta \in [-\pi,\pi)$ with $V=
\tan(\theta/2)$, so that $\infty$ corresponds to $\pm \pi$ and the
bifurcation correspond to $\theta =0$. Three independent vector
fields generating the full action $PSL(2,\bR)$ group on the extended
manifold $\bS\times \Sigma$ are \beq {\cal D} = V{\partial_V} = \sin
\theta \partial_\theta\:,\:\:\: {\cal K} = \frac{2}{1+V^2}\partial_V
= \partial_\theta\:,\:\:\: {\cal H}= \partial_V = (1+\cos \theta)
\partial_\theta\:. \label{vf} \eeq Integrating the transformations
generated by linear combinations of these vectors one obtains the
action of any $g\in PSL(2,\bR)$ on $\bS^1\times \Sigma$. $g$
transforms $p\in \bS^1\times \Sigma$ to the point $g(p) \in
\bS^1\times \Sigma$. (See \cite{MP2,MP4} for the explicit expression
of $g(p)$). Finally, the action of $PSL(2,\bR)$ on $\bS^1\times
\Sigma$ induces an active action \linebreak on fields: \beq \hat{\phi}_\zeta(p)
\mapsto \hat{\phi}_\zeta(g^{-1}(p))\:,\quad \mbox{for every $g\in
PSL(2,\bR)$}\:,\label{action} \eeq which preserves commutation
relations. This is valid for any value of $\zeta$, including
$\zeta=0$. Notice that all this structure is quite universal: the
vector fields $\cD, \cK,\cH$ do not depend on the state $\Psi_\zeta$
characterizing the mass of the black hole, but they depend only on
the choice of the preferred coordinate $V$, that is $\Upsilon$.

{\em If $\zeta =0$}, it is possible to {\em unitarily implement}
that action (\ref{action}) of $PSL(2,\bR)$ on $\hat\phi$; in other
words \cite{MP6,MP66}, there is a (strongly continuous) unitary
representation $U$ of  $PSL(2,\bR)$ such that  $$U_g \hat \phi(p)
U^{\dagger}_g = \hat{\phi}(g^{-1}(p))\:,
 \quad \mbox{for every $g\in PSL(2,\bR)$}\:.$$
Furthermore,   it turns out that the state $\Psi$ is {\em invariant
under  $U$} itself, that is $$U_g \Psi = \Psi\:,\quad \mbox{for
every $g\in PSL(2,\bR)$}\:. $$ To define $U$ one introduces the {\em
stress tensor} $$\hat{T}(V,s) \ = \:\: :\spa\partial_V \hat\phi
\partial_V \hat\phi\spa: (V,s)\:.$$ The state $\Psi$ enters the definition by the normal
ordering prescription it being  defined by subtracting $\langle\Psi
|\hat{\phi} (V',s')\hat{\phi} (V,s) \Psi \rangle$ before applying
derivatives and then smoothing with a product of delta in $V,V'$ and
$s,s'$. One can smear $\hat T$ with a vector field $\cX:=
X(V,s)\partial_V$ obtaining the operator
$$T[\cX] = \int_\bF X(V,s) \hat{T}(V,s) dV \wedge \omega_\Sigma\:.$$
It is possible to show \cite{MP6,MP66} that the three operators, obtained by smearing $\hat T$ with $\cD,\cK,\cH$ respectively,
\begin{eqnarray}
T[\cD] &=& \frac{1}{4i}\sum_{j, k>0} :\spa J_{-k}^{(j)}J_{k+1}^{(j)}\spa: -
:\spa J_{-k}^{(j)}J_{k-1}^{(j)}\spa:\:,\\
 T[\cK] &=&
\frac{1}{2}\sum_{k\in \bZ, j\in \bN} \spa: J_{-k}^{(j)}J_{k}^{(j)}\spa :\:,\\
T[\cH] &=& \frac{1}{4}\sum_{k\in \bZ, j\in \bN} 2:\spa
J_{-k}^{(j)}J_{k}^{(j)}\spa : + :\spa J_{-k}^{(j)}J_{k+1}^{(j)} \spa
: + :\spa J_{-k}^{(j)}J_{k-1}^{(j)} \spa :\:,
\label{Hvectors}\end{eqnarray} are the very generators of the
unitary representation $U$ of $PSL(2,\bR)$ which implements the  action of
$PSL(2,\bR)$ on  $\hat{\phi}(V,s)$ leaving fixed $\Psi$. They, in
fact,  generate the one-parameter subgroups of $U$  associated with
the diffeomorphisms due to vector fields $\cD,\cK,\cH$ respectively.
The fact that
$T[\cD],T[\cK], T[\cH]$ enjoy correct commutation relations is not
enough to prove the existence of the unitary representation. Rather
the existence is consequence  of the presence of an invariant and
dense space of analytic vectors for $T[\cD]^2+T[\cK]^2+T[\cH]^2$ and
know nontrivial theorems by Nelson. See \cite{MP1,MP2,MP4,MP6,MP66} for
details and references.

The normal ordering prescription for operators $P_n$ is defined by
that $:\spa P_kP_h\spa: = P_hP_k$ if  $h<0$ and $k>0$, or $:\spa
P_kP_h \spa: = P_kP_h$ otherwise.

All that is mathematically interesting, but it is unsatisfactory
form a physical point of view if we want to describe classical
geometric properties of the horizon as consequences of quantum
properties. Indeed, in this way, the quantum picture admits a too
large unitary symmetry group which exists anyway, no matter if the
manifold is extended by adding the points at infinity or not. This
larger group does not correspond to the geometrical shape of the
physical manifold $\bF$: The transformations associated with vector
fields $\cK$ do not preserve the physical manifold $\bF$, they move
some points in the physical manifold to infinity. The
transformations associated with vector field $\cH$ transforms $\bF$
into $\bF$ itself but they encompass translations of the bifurcation
surface $\Sigma$ which we have assumed to be fixed at the beginning.
Only the vector $\cD$ may have a completely satisfactory physical
meaning as  it simply generates dilatations of the coordinate $V$
transforming $\bF$ into $\bF$ itself and leaving fixed $\Sigma$. One
expects that there is some way, at quantum level, to get rid of the
physically irrelevant symmetries and that the unphysical symmetries
are removed from the scenario once one has fixed the quantum state
of a black hole. In fact this is the case. Switching on $\zeta\neq
0$ the situation changes dramatically and  one gets  automatically
rid of the unphysical transformations picking out the physical ones.
Indeed, the following result can  be proved (it is a stronger version than Theorem 3.2 \cite{MP6,MP66}). \\

\noindent \noindent {\bf Theorem 2}. {\em If $\zeta\neq 0$, there is
no unitary representation of the whole group $PSL(2,\bR)$ which
unitarily implements the  action of $PSL(2,\bR)$ on the field
$\hat{\phi}_\zeta$ (\ref{cl-qu}) referred to  $\Psi_\zeta$. Only the
subgroup associated with ${\cal D}$ admits unitary implementation
\beq e^{-i\tau H_\zeta}\hat\phi_\zeta(V,s)e^{i\tau H_\zeta} = \hat
\phi_\zeta(e^{-\tau}V,s)\label{HU} \eeq and $\Psi_\zeta$ is
invariant under that unitary representation of the group \beq
e^{-i\tau H_\zeta}\Psi_\zeta = \Psi_\zeta\:. \eeq} \vspace{-0.2cm}
\noindent {\bf Sketch of Proof}. ~If $\zeta \in \bR$ is fixed
arbitrarily,  and $\omega$ varies in the class of the admissible
real forms used to smear the field operator, the class of all of
unitary operators in the Fock space based on $\Psi_\zeta$
$$W_\zeta(\omega) = \exp\left\{i\int_\bF \hat{\phi} \omega\right\} =
\exp\left\{i\int_\bF \hat{\phi}_0 \omega + \zeta \ln|V|
\omega\right\}$$ turns out to be irreducible (see \cite{MP6,MP66}). If
$g\in PSL(2,\bR)$, $W_\zeta(\omega) \mapsto W_\zeta(\omega^{(g)})$
denotes the geometric action of the group on the operators
$W_\zeta(\omega)$. We know that, for $\zeta=0$, this action can be
unitarily implemented (Theorem 3.2 in \cite{MP6,MP66}). This is
equivalent to say that  there is a unitary representation $U$ of
$PSL(2,\bR)$ such that \beq U_g \exp\left\{i\int_\bF \hat{\phi}_0
\omega\right\}U^\dagger_g = \exp\left\{i\int_\bF \hat{\phi}_0
\omega^{(g)}\right\}\:.\label{A}\eeq For that representation it
holds $U_g \Psi_\zeta =\Psi_\zeta$ Suppose now that the action can
be implemented for $\zeta\neq 0$ by means of the unitary
representation of $PSL(2,\bR)$, $V^{(\zeta)}$. In other words \beq
V_g^{(\zeta)}\exp\left\{i\int_\bF \hat{\phi}_0 \omega + \zeta \ln|V|
\omega\right\} V_g^{(\zeta)\dagger} =
 \exp\left\{ i\int_\bF \hat{\phi}_0 \omega^{(g)}\right\}  \exp\left\{i\int_\bF\zeta \ln|V| \omega^{(g)}\right\}\:.\label{B}\eeq
Consider the unitary operator $S_g= U_g^\dagger V_g^{(\zeta)}$. Due
to (\ref{A}) and (\ref{B}), one simply gets \beq S_g W_0(\omega) =
e^{ic_{g,\omega}}W_0(\omega)S_g\:,\label{PG}\eeq where
$c_{g,\omega}$ is the real  $\zeta \int_{\bF} [\ln|V| (\omega^{(g)}
- \omega)] $. From standard manipulations working with the spectral
measure of $S_g$ one finds that (\ref{PG}) implies, if $P_E$ is any
projector in the spectral measure of $S_g$:
$$ P_E W_0(\omega) = e^{ic_{g,\omega}}W_0(\omega)P_E\:.\label{PG3}$$
Since the spectral measure is complete and $W_0(\omega)\neq 0$,
there must be some projector $P_E$ such that  $P_E W_0(\omega) \neq
0$  and $W_0(\omega)P_E\neq 0$. For for all those projectors the
identity above is possible only for $c_{g,\omega} = 0$. Therefore
every projection space (including those whose projectors do not
satisfy $P_E W_0(\omega) \neq 0$  and $W_0(\omega)P_E\neq 0$) turns
out to be invariant with respect to  $W_0(\omega)$. The result is
valid for every $W_0(\omega)$. This is impossible (since the
considered operator form an irreducible class as said at the
beginning) unless $S_g = e^{ia_g} I$ for some real $a_g$. In other
words: $V_g = e^{ia_g}U_g$. Inserting it in (\ref{B}) and comparing
with (\ref{A}) one finds that the  constraints $c_{g,\omega} = 0$
must hold true, that is
$$\int_{\bF} [\ln|V| (\omega^{(g)} - \omega)] = 0$$
for every $g\in PSL(2,\bR)$ and every smearing form $\omega$. It has
been established in the proof of Theorem 4.1 of \cite{MP6,MP66} that this
is possible if and only if $g$ belongs to the one-parameter subgroup
of $PSL(2,\bR)$ generated by $\cD$. The unitary representation of
that subgroup has been constructed explicitly finding (\ref{HU}) and
(\ref{H0}). Moreover, in the same theorem, it has been similarly
proved that $\Psi_\zeta$ is invariant under the action of  that
unitary representation.
These results conclude the proof.  $\Box$\\

The thesis shows that the strongest notion of {\em spontaneously
breaking of ($PSL(2,\bR)$) symmetry} used in algebraic quantum field
theory arises: There is a group of transformations (automorphisms),
in our case associated with $PSL(2,\bR)$, of the algebra of the
fields which cannot be completely implemented unitarily. The
self-adjoint generator $H_\zeta$ of the surviving group of symmetry
turns out to be \cite{MP6,MP66}: \beq H_\zeta  =  \int_{\bF}
V\::\spa\partial_V\hat{\phi}_0 \partial_V\hat{\phi}_0\spa :(V,s) \:
dV \wedge \omega_\Sigma(s) \label{H0}\:, \eeq the normal ordering
prescription  being  defined by subtracting $\langle\Psi_\zeta
|\hat{\phi}_0 (V',s')\hat{\phi}_0 (V,s) \Psi_\zeta \rangle$ before
applying derivatives (which is equivalent to subtract
$\langle\Psi|\hat{\phi} (V',s')\hat{\phi} (V,s) \Psi \rangle$). This
definition is equivalent to that expected by formal calculus: \beq
H_\zeta  =   T_\zeta[\cD] = \int_{\bF} V\: \hat{T}_\zeta(V,s)\: dV
\wedge \omega_\Sigma(s) \label{H}\:,\eeq where
$$\hat{T}_\zeta(V,s)= :\spa\partial_V\hat{\phi}_\zeta \partial_V\hat{\phi}_\zeta\spa:  (V,s)\:,$$
{\em with the above-defined notion of normal ordering},
assuming linearity and $:\spa\hat{\phi}_0\spa: = \hat{\phi}_0$.
Indeed, let $v$ be the  parameter of the integral curves of $\cD$, so that $v= \ln |V|$ and $v\in \bR,s\in \Sigma$ define
a coordinate system on both $\bF_>$
and $\bF_<$ separately.
 Starting from (\ref{H}), one has:
$$ H_\zeta = \lim_{N\to +\infty}  \left\{\int_{\bF_>}   \chi_N(v)
:\spa\frac{\partial\hat{\phi}_0}{\partial v}
\frac{\partial\hat{\phi}_0}{\partial v} \spa: \spa (V_+(v),s) \:dv\wedge\omega_\Sigma(s) +
\zeta^2 A_0 \int_{\bR}   \chi_N(v) dv\right. $$
\beq - \left.\int_{\bF_<}   \chi_N(v) :\spa
\frac{\partial\hat{\phi}_0}{\partial v}\frac{\partial\hat{\phi}_0}{\partial v}\spa: \spa(V_-(v),s)\:
dv\wedge\omega_\Sigma(s) -
 \zeta^2 A_0 \int_{\bR}   \chi_N(v) dv\right\}\:,\label{big}\eeq
where from now on $A_0 = 4\pi r_0^2$. Moreover $\chi_N(v)$ is a
smooth function with compact support in the interior of $F_<$ and
$F_>$ separately, which tends to the constant function $1$ for $N\to
+\infty$ and $V_\pm(v) = \pm e^v$. We have omitted a term in each
line proportional to $(\partial_v \chi_N)\phi_0$ (using derivation
by parts).  Those terms on, respectively,  $\bF_<$ and $\bF_>$  give
no contribution separately as $N\to \infty$ with our hypotheses on
$\chi_N$. The remaining two constant terms at the end of each line
in brackets cancel out each other and this computation shows that
(\ref{H}) is  equivalent to (\ref{H0}).

Physically speaking, with the given definition of $\zeta>0$,
$\zeta^{-1}\cD$ is just the restriction to $\bF$ of the Killing
vector of the spacetime defining the static time of the external
region of black holes. If, as above, $v$ is the parameter of
integral curves of $\cD$, $\zeta v$ itself is the limit of Killing
time towards $\bF$. At space infinity this notion of time coincides
with Minkowski time. Let us restrict the algebra of observables
associated with the field $\hat\phi$ to the region $\bF_>$ where
$\cD$ is future directed. This is done by smearing the fields with
forms  completely supported in $\bF_>$. Therein one can adopt
coordinates $v,s$ as above  obtaining:
\begin{eqnarray}
\langle \Psi_\zeta |\partial_v\hat {\phi}_\zeta(v,s) \Psi_\zeta\rangle &=& \zeta \:,\label{former'}\\
\langle \Psi_\zeta |\partial_v \hat {\phi}_\zeta(v,s) \partial_{v'}\hat{\phi}_\zeta(v',s')\Psi_\zeta\rangle &=& -\frac{\delta(s,s')}{4\pi}
\frac{e^{v-v'}}{(1-e^{v-v'})^2}
 \:,\label{latter'}
\end{eqnarray}
Take the above-mentioned smearing procedure into account and the
fact that one-point and two-point functions reconstruct all
$n$-point functions class as well.  Therefore, from (\ref{former'})
and (\ref{latter'}), it follows that that the $n$-point functions
are invariant under $\cD$ displacements. Furthermore, performing
Wick rotation $v\to iv$, one obtains $2\pi$ periodicity in the
variable $v$. This is nothing but the analytic version of well-known
{\em KMS condition} \cite{Haag,BR,BR2}. These fact can be
summarized as:\\

\noindent {\bf Theorem 3}. {\em Every state $\Psi_\zeta$ (including
$\zeta=0$), restricted to the algebra of observables localized at
$\bF_>$, is invariant under the transformations generated by $\cD=
\partial_v$ and it is furthermore thermal with respect to the time
$v$ with inverse temperature $\beta = 2\pi$. As a consequence,
adopting the physical ``time coordinate'' $\zeta v$ which accounts
for the actual size of the Black hole (enclosed  in the parameter
$\zeta$), the inverse temperature $\beta$
turns out to be just Hawking's value $\beta_H = 8\pi M$.}\\

It is furthermore possible to argue  that the state $\Psi_\zeta$
contains a Bose-Einstein condensate of quanta with respect to the
generator of $v$ displacements for the theory restricted to $\bF_>$.
We have provided different reasons for this conclusion in
\cite{MP6,MP66}. In particular the non-vanishing one-point function
(\ref{former}) is a typical phenomenon in Bose-Einstein condensation
(see chapter 6 of \cite{Popov}). The decomposition (\ref{cl-qu}) of
the field operator into a ``quantum'' $\hat{\phi}_0(v,s)$ part (with
vanishing expectation value) and a ``classical'', {\em i.e.}
commuting with all the elements of the algebra, part $\zeta vI$, is
typical of the theoretical description of a boson system containing
a Bose-Einstein condensate; the classical part  $\zeta v = \langle
\Psi_\zeta |\hat \phi(v,s) \Psi_\zeta\rangle $ plays the role of a
{\em order  parameter} \cite{DGPS,Popov}. The classical part is
responsible for the macroscopic properties of the state. Considering
separately the two disjoint regions of $\bF$, $\bF_<$ and $\bF_>$
and looking again at (\ref{big}), $H_\zeta$  is recognized to be
made of two contributions $H_\zeta^{(<)}$, $H_\zeta^{(>)}$
respectively localized at $\bF_<$ and $\bF_>$. The two terms have
opposite signs corresponding to the fact that the Killing vector
$\partial_{v}$ changes orientation passing from $\bF_<$ to $\bF_>$.
As \beq H^{(>)}_\zeta =  \int_{\bF_>} V T_\zeta(V,s)
\:dV\wedge\omega_\Sigma(s) + \zeta^2 A_0 \int_{\bR}  dv
\label{big2}\eeq it contains the {\em classical} volume-divergent
term
$$\langle \Psi_\zeta |H^{(>)}_\zeta \Psi_\zeta \rangle =  \zeta^2 A_0\int_{\bR} dv\:.$$
This can be interpreted as the ``macroscopic energy'', with respect
to the Hamiltonian $H^{(>)}_\zeta$, due to the  Bose-Einstein
condensate localized at $\bF_>$, whose density is {\em finite} and
amounts to $\zeta^2 A_0$.

As a final comment we stress that, in \cite{MP6,MP66}, we have proved
that any  state  $\Psi_{\zeta}$ defines an  {\em extremal state} in
the convex set of KMS states on the $C^*$-algebra of Weyl observable
defined on  $\bF_>$ at inverse  temperature $2\pi$ with respect to
$\partial_v$ and that different choices of $\zeta$ individuate not
unitarily equivalent representations. The usual  interpretation of
this couple of results is that the states  $\Psi_\zeta$, restricted
to the observables in the physical region $\bF_>$, coincide with
{\em different thermodynamical  phases} of the same system at the
temperature $2\pi$ (see V.1.5 in \cite{Haag}).

\subsection{Properties of $\Upsilon$ and $\hat\rho$: Feigin-Fuchs Stress Tensor}
Let us consider the realization of CCR for the field $\hat \rho$ in
the Fock representation based on the vacuum vector $\Upsilon$ which
singles out the preferred admissible null coordinate $V$. In this
case there is no spontaneous breaking of symmetry. However, due to
the particular affine transformation rule  (\ref{affinerhoF}) of the
field $\hat\rho$, there are anyway some analogies with the CCR
realization for the field $\hat \phi$ referred to the state
$\Psi_\zeta$. Using the coordinate patches $(v, s)$ on $\bF_+$ with
$\partial_v =  \cD$ and exploiting (\ref{affinerhoF}), the field
takes the form \beq \hat \rho(v, s) = \hat{\rho}(V(v), s) +
\ln|V|\:I\:. \eeq This equation resembles (\ref{cl-qu}) with
$\zeta=1$ and thus one finds in particular:
\begin{eqnarray}
\langle \Upsilon |\partial_v\hat \rho(v,s) \Upsilon \rangle &=& 1 \:,\label{former2'}\\
\langle \Upsilon |\partial_v \hat \rho(v,s) \partial_{v'}\hat \rho(v',s')\Upsilon\rangle &=& -\frac{\delta(s,s')}{4\pi}
\frac{e^{v-v'}}{(1-e^{v-v'})^2}
 \:.\label{latter2'}
\end{eqnarray}
As a consequence, analogous comments on the interplay of state $\Upsilon$
and the algebra of fields $\rho(v, s)$ (notice that they are defined in the region $\bF_>$) may be stated.
In particular:\\

\noindent {\bf Theorem 4}. {\em The state $\Upsilon$ restricted to
the algebra of observables localized at $\bF_>$ turns out to be a
thermal (KMS) state
 with respect to $\partial_{\zeta v}$
at Hawking temperature.}\\

We want now to focus on the stress tensor generating the action of
$SL(2,\bR)$ on the considered affine field. Fix an admissible global
null coordinate frame inducing  coordinates $(x^+,s)$ on $\bF$. A
stress tensor, called {\em Feigin-Fuchs stress tensor} \cite{FMS},
can be defined as follows. \beq\label{st} \hat \cT(x^+,s)= \:
:\spa\pa_{x^+}\hat{\rho}\pa_{x^+}\hat{\rho} \spa:(x^+,s) -2\alpha
\pa_{x^+}\pa_{x^+}\hat{\rho}_{x^+}(x^+,s)\:. \eeq The normal ordered
product  with respect to $\Upsilon$,
$:\spa\pa_{x^+}\hat{\rho}\pa_{x^+}\hat{\rho}\spa :(x^+,s)$, is
defined by taking the limit for $(x,z) \to (x^+,s)$ of
$:\spa\pa_{x^+}\hat{\rho}(x,z)\pa_{x^+}\hat{\rho}(x^+,s)\spa:$, the
latter being defined as
$$ \pa_{x^+}\hat{\rho}(x,z)
\pa_{x^+}\hat{\rho}(x^+,s)
- \langle\Upsilon|\pa_{x^+}\hat{\rho}(x,z)\pa_{x^+}\hat{\rho}(x^+,s)\Upsilon \rangle
 + \langle\Upsilon |\pa_{x^+}\hat{\rho}(x,z)\Upsilon \rangle \langle\Upsilon |\pa_{x^+}\hat{\rho}(x^+,s)\Upsilon \rangle \:.$$
The last term, which vanishes with our choice for $\Upsilon$ if working in coordinates $V,s$,
is necessary in the general case to
reproduce correct affine transformations for
$:\spa\pa_{x^+}\hat{\rho}(x,z)\pa_{x^+}\hat{\rho}(x^+,s)\spa:$ under changes of coordinates for each field in the product
separately. The stress tensor can  be smeared with vector fields ${\cal X}= X(x^+,s)\partial_{x^+}$:
$$
\cT[{\cal X}]=\int_\bF  X(x^+,s)\: \hat{\cT}(x^+,s)
\:dx^+\wedge \omega_\Sigma\:.
$$
As a consequence one obtains (where it is understood that the fields
are smeared with forms as usual) \beq
\de\hat{\rho}(x^+,s)=-i\left[\cT[{\cal
X}],\:\hat{\rho}(x^+,s)\right]= X(x^+,s)\pa_{x^+}\hat{\rho}(x^+,s)+
\alpha \pa_{x^+} X(x^+,s) \label{cq1}\:, \eeq
which is nothing but the infinitesimal version of transformation (\ref{affinerhoF}) provided $\alpha =1$.\\
It is worth to investigate whether or not  $\cT[{\cal D}]$, $\cT[{\cal K}]$, $\cT[{\cal H}]$ are the self-adjoint
 generators of a unitary representation
which implements the active action of $PSL(2,\bR)$ on the field $\hat{\rho}(V,s)$:
$$\hat{\rho}(V,s) \mapsto \hat{\rho}(g^{-1}(V),s) + \ln \frac{dg^{-1}(V)}{dV}\:, \quad \mbox{for any $g\in PSL(2,\bR)$.}$$ \newpage
\noindent The answer is interesting: once again spontaneous breaking of $PSL(2,\bR)$
symmetry arises but now the surviving subgroup is larger than the
analog for $\hat\phi$. Indeed, the following set of result can be
proved with dealing with similarly to the proof of Theorem 2.\\

\noindent {\bf Theorem 5}. {\em Working in coordinates $V,s$ and referring to the representation of
$\hat\rho$ based on $\Upsilon$:\\
(1) there is no unitary representation of $PSL(2,\bR)$ which implements
the action of the whole group $PSL(2,\bR)$ on the field $\hat{\rho}$.\\
(2) There is anyway a (strongly continuous) unitary representation
$U^{(\Delta)}$ of the $2$-dimensional subgroup $\Delta$ of
$PSL(2,\bR)$ generated by $\cD$ and $\cH$ together,
which implements the  action of $\Delta$ on the field
$\hat{\rho}(V,s)$.
$$U^{(\Delta)}_g\hat{\rho}(V,s)U^{(\Delta)\dagger}_g =
\hat{\rho}(g^{-1}(V),s) + \ln \frac{dg^{-1}(V)}{dV}\:, \quad \mbox{for any $g\in \Delta$.}$$
The self-adjoint generators
of $U^{(\Delta)}$  are $\cT[{\cal D}]$ and $\cT[{\cal H}]$ (with $\alpha =1$).\\
(4)  $\Upsilon$ is invariant under $U^{(\Delta)}$.}\\

Notice
that $\cD$ and $\cH$ form a sub Lie algebra of that of $PSL(2,\bR)$,
whereas the remaining couples in the triple $\cD,\cH,\cK$ do not.
The explicit form of the generators $\cT[{\cal D}]$ and $\cT[{\cal
H}]$ can be obtained in function of the operators $P^{(j)}_k$. With
the same definition of normal ordering for those operators as that
given for operators $J^{(j)}_n$, one has:
\begin{eqnarray}
\cT[{\cal D}] &= & \frac{1}{4i} \sum_{n\in \bZ, j\in \bN} :\spa P^{(j)}_{-n}P^{(j)}_{n+1}\spa: - :\spa P^{(j)}_{-n}P^{(j)}_{n-1}\spa:\:,\\
\cT[{\cal H}] &=& \frac{1}{4} \sum_{n\in \bZ, j\in \bN} :\spa P^{(j)}_{-n}P^{(j)}_{n+1}\spa: + :\spa P^{(j)}_{-n}P^{(j)}_{n-1}\spa: +2
:\spa P^{(j)}_{-n}P^{(j)}_{n}\spa:\:.
\end{eqnarray}

Dropping the dependence on $s$, ${\bf T}(V,s)$ defined in
(\ref{st}) is the stress tensor of a $1$-dimensional {\em Coulomb
gas} \cite{FMS}.  As is well known it does not transform as a
tensor: By direct inspection one finds that, under changes of
coordinates $x^+\to x'^+$, \beq \hat{\cT}(x'^+,s)=\at\frac{\pa
x^+}{\pa x'^+}\ct^2\, \hat{\cT}(x^+,s) -2\alpha^2\ag x^+,x'^+\cg \:,
\label{schwarz} \eeq where $\{z,x\}$ is the Schwarzian derivative
(which vanishes if $x^+\to x'^+$ is a transformation in
$PSL(2,\bR)$)
$$\{z,x\}=\frac{\frac{d^3z}{dx^3}}{\frac{dz}{dx}}-\frac{3}{2}\left(\frac{\frac{d^2z}{dx^2}}{\frac{dz}{dx}}\right)^2\:.$$
The coefficient in front of the Schwarzian derivative in
(\ref{schwarz}) differs from that found in the literature (e.g. see
\cite{FMS}) also because here we use a normal ordering procedure
referred to unique reference state, $\Upsilon$, for all coordinate
frames. We stress that, anyway,  $\Upsilon$ is the vacuum state only
for coordinate $V,s$. Let us restrict ourselves to $\bF_>$ and use
coordinate $v$ with $\partial_v = \cD$ therein. If $x^+=V$ and
$x'^+= v$ one \linebreak finds by (\ref{schwarz}) \beq \hat{\cT}(v,s) =
\at\frac{\pa V}{\pa v}\ct^2 \hat{\cT}(V,s) + 1 \:.\label{TT}\eeq The
formally self-adjoint generator for  the field $\hat{\rho}(v,s)$,
defined on $\bF_>$ and generating the transformations associated
with the vector field $\cD = V\partial_V = \partial_v$, is
$${\bf H}^{(>)} = \int_{\bF_>} \hat{\cT}(v,s) dv\wedge \omega_\Sigma\:.$$
 From (\ref{TT}) one finds:
\beq {\bf H}^{(>)} = \int_{\bF_>} V \:\hat{\cT}(V,s) \:dV\wedge
\omega_\Sigma + \: 1\: A_0\int_\bR dv\:,\eeq This formula strongly
resembles (\ref{big2}) for $\zeta=1$  also if it has been obtained,
mathematically speaking, by a completely different way and using the
field $\hat\rho$ with property of transformations very different
than those of the scalar $\hat{\phi}_\zeta$.

\section{Thermodynamical Quantities: Free Energy and Entropy}
Assume that the states $\Upsilon$ and $\Psi_\zeta$ are given and let
$v_\zeta = \zeta v$,  $v$ being the integral parameter of $\cD$ on
$\bF_>$. If $\phi^+$ denotes the classical field restricted to
$\bF$, one has  $\phi^+(v) = \langle \hat{\phi}_\zeta(v) \rangle =
v_\zeta$ and this is in agreement with the fact that
$\partial_{\phi^+} - \partial_{\phi^-}$ is the Killing field
defining Schwarzschild time in spacetime (see section \ref{sec1}).
The temperature of the state $\Psi_\zeta$ coincides with Hawking one
when referring to the ``time'' $v_\zeta$. Therefore let us focus
attention on the  generator of $v_\zeta$ displacements $\zeta^{-1}
H_\zeta^{(>)}$ whose ``density of energy'', due to the condensate,
is $$\langle \Psi_\zeta | \zeta^{-1} H_\zeta^{(>)} \Psi_\zeta
\rangle /\int_\bR dv = \zeta A_0\:.$$ We try to give some physical
interpretation to that density of energy. First of all notice that
we are considering a system containing Bose-Einstein condensate at
temperature $\beta_H^{-1}>0$. This picture has to be discussed in
the approach of grand canonical ensemble in the thermodynamical
limit  and the chemical potential $\mu$ must vanish in this
situation. In this context the generator $\zeta^{-1} H_\zeta^{(>)}$
which generate the one-parameter group of transformations verifying
KMS conditions is that of a grand canonical ensemble and its
averaged value has to be interpreted as the density of {\em free
energy} of the system (see chapter V of \cite{Haag}) rather than its
energy.  Notice that the density is computed with respect to the
parameter $v$ which is universal, not depending on $\zeta$, and
valid for every black hole. We recall the reader that  $\beta_H =
8\pi M$ and $\zeta= 4M$. We conclude that $$F(\beta_H) :=
\zeta(\beta_H)A_0$$ is a density of free energy. Concerning the
densities of energy and entropy one has: \beq E = \frac{\pa}{\pa
\be_H} \be_H F\: ,\qquad  S=\be_H^2\frac{\pa}{\pa \be_H} F\:.
\label{last} \eeq where  some terms in the right hand side have been
dropped because they are proportional to $\mu=0$. For the case
$n=4$, fixing the universal parameter $r_0$ as
 $r_0 = 1/(4\sqrt{\pi})$ one gets, if $M$ is the mass of the black hole and $A$ the area of its horizon:
$$ F=\frac{M}{2} \: ,\qquad  E=M \:,\qquad  S={4\pi M^2}=\frac{A}{4}\:.$$

\section{Final Comments and Open Issues}
Even if we have focused  on the $S$-wave sector, which captures just
the radial modes of the metric leaving outside, for instance,
gravitational  waves, the results presented  in (\ref{last})  are
suggestive and one may hardly think that they arise by chance. There
are anyway two problems to tackle in order to be confident in  our
approach to understand black hole thermodynamics from a quantum
point of view. First of all the parameter $r_0$ is universal but
there is no way to fix it at the beginning, within our approach.
However it remains that the densities of energy and entropy scale as
the energy and the entropy of black holes modulo $r_0$ which does
not depend on the size of the black hole. The second point concerns
the fact that $E$ and $S$ are {\em densities} of energy and entropy,
but they are compared with energy and entropy of black holes. These
densities are evaluated with respect to an universal---and
dimensionless if introducing dimensions---parameter $v$, which is
proper of the arena where to represent all different black holes
(each depending on its own  value $\zeta$ of the order parameter
used to break the conformal symmetry). Notice also that the
densities are referred to observables {\em homogeneously} spread
along the Killing horizon, that is the evolution in time of the
$2$-sphere defining the horizon of the black hole at fixed time.  A
Cauchy surface for the whole Kruskal spacetime intersect, at every
time, the Killing horizon in such a $2$-sphere (not necessarily the
same). Usually handled quantities of black holes  are referred to
that $2$-sphere. If a relation exists between those two classes of
quantities (spread on the whole horizon or defined on the
$2$-sphere) it is reasonable that quantities  defined on the $2D$
sections of $\bF$ are the densities of the corresponding ones
homogeneously spread along $\bF$. However this issue deserves
further investigation.

Further investigation are also necessary to translate horizon
quantization proposed here to that presumably existing in the bulk.
If this task seems to be straightforward regarding the field $\phi$,
it seems to be very difficult for the field $\rho$ due to Einstein
equations. To this end it is worthwhile stressing that, in the
$3$-dimensional case, $\rho$ is a Liouville field in the bulk whose
quantization is not simple at all. In this case it seems that the
horizon fields $\rho_\bF$ could play the role of a chiral current
emerging from  canonical quantization of the Liouville fields. In
the general case the situation is also more complicated because of
the presence of the field $\eta$.  It enters the equation of motion
of  $\rho$, so that, quantization of $\eta$ needs to be considered
as well.

In this paper, we have considered the field $\rho$ and $\phi$ as
almost independent. Actually, on the horizon the following classical
equation for classical fields holds: \beq
\partial_+ \rho = \zeta^{-1}\partial_+\phi^+&+&\frac{\partial^2_+ \phi^+}{\partial_+ \phi^+}\:,\label{fin}\\
\frac{1}{2}\cV(\eta_C) &=& \zeta^{-1} \label{fin2}\:. \eeq These
relations are nothing but the Einstein equation on the horizon. The
requirement $\phi^+= \zeta v$ is nothing but a solution of that
equation in suitable coordinates. We have considered it as a
relation valid for the expectation value of the field. A posteriori
(\ref{fin}) and (\ref{fin2}) have to be considered as a kind of
thermodynamical relations. Their meaning or, more appropriately, the
corresponding equations at quantum level controlling the fields
$\hat{\rho}$ and $\hat{\phi}$ are not yet understood.

As a final comment we notice that $\hat{\phi}$ may be viewed  as a
non-commutative light-coordinate on the horizon, in fact on the
state implementing symmetry breaking the expectation value of
$\media{\hat{\rho}(v)}=\zeta v$ defines a preferred coordinate
$\zeta v$ on the horizon. This issue deserves further investigation.

To conclude we want to make a technical comment
concerning the possibility of constructing a unitarizable
representation of Virasoro algebra using the Feigin-Fuchs  stress
tensor (\ref{st}) and smearing it with the complex vector fields
\beq \cL_n= i\, e^{in\theta}\,\frac{\pa}{\pa \theta}\:. \eeq Here we
have added the point at infinity to the light geodesics of $\bF$
obtaining the extended (unphysical) manifold $\bS^1 \times \Sigma$.
If $\bS^1 = [-\pi, \pi)$ with $-\pi\equiv \pi$ and $\theta$ ranges
in $\bS^1$, $V=\tan(\theta/2)$. The fields $\{\cL_n\}_{n\in \bZ}$
enjoy Virasoro commutation relations without central charge and
satisfy Hermiticity condition with respect to the involution
$\imath(X) = -\overline{X}$, respectively \beq \{\cL_n,\cL_m\}=
(n-m) \cL_{n+m}\:, \:\:\:\:\: \mbox{$\imath(\cL_n)= \cL_{-n}$,} \eeq
$\{\:\cdot,\:\cdot\}$ denoting the usual Lie bracket (see \cite{Kac}
and Section III of \cite{MP6,MP66} for further details). However, a direct
computation shows that, if $\cT[\cL_n]$ is that refereed to the
preferred coordinates $V,s$ defining \linebreak the vacuum
$\Upsilon$, \beq\label{genvir} \cT[\cL_n] = \frac{1}{2}\sum_{k\in
\bZ, j\in \bN}P^{(j)}_{-k}P^{(j)}_{k+n} -i\alpha \sqrt{4\pi A_0}
\left[n P^{(0)}_{n} + \sum_{k\in \bZ} (-1)^k \sigma_k
P^{(0)}_{k+n}\right] \eeq where $j=0$ individuates the constant
function among the orthonormal complete set $\{u_j\}_{j\in \bN}$,
$\sigma_0=0$ otherwise $\sigma_k$ is the sign of $k\in \bZ$. It is
simply proved that, in general,  the commutator of $\cT[\cL_n]$ and
$\cT[\cL_m]$ produces an infinite constant term among other
operatorial terms. In general it is possible to cancel out difficult
terms using suitable linear combinations of operators $\cT[\cL_n]$,
in particular those corresponding to $\cT[\cD]$ and $\cT[\cH]$. The
reason is that not all diffeomorphisms of the circle preserve the
physical manifold $\bF_+$. Only those which do it can be represented
by means of $\cT$.

\section*{Acknowledgements}
The work of N.P. is supported in part by the ERC Advanced Grant 227458 OACFT ``Operator Algebras and Conformal Field Theory."


\bibliographystyle{mdpi}
\makeatletter
\renewcommand\@biblabel[1]{#1. }
\makeatother

\end{document}